\documentclass[sigconf]{acmart}
\usepackage{xcolor}
\usepackage{colortbl}
\usepackage{tcolorbox}
\usepackage{listings}
\usepackage{makecell}
\usepackage{multirow}
\usepackage{amsmath,amsfonts}

\usepackage{amssymb}
\usepackage{hyperref} 
\usepackage{algorithm}
\usepackage{algorithmic}
\usepackage{graphicx}
\usepackage{fbox}
\usepackage{textcomp}
\usepackage{minibox}
\usepackage{balance}
\usepackage{enumitem}
\usepackage{xspace}
\usepackage[T1]{fontenc}
\usepackage{booktabs}
\usepackage{threeparttable}
\usepackage[]{collab}
\usepackage{url}
\usepackage{natbib}

\usepackage[utf8]{inputenc}
\usepackage{color}
\usepackage{calligra}
\usepackage[normalem]{ulem}
\usepackage{indentfirst}
\usepackage{nicematrix}
\usepackage{fontawesome5}
\usepackage{pifont}

\definecolor{ashgrey}{rgb}{0.98, 0.91, 0.71}
\definecolor{grey}{rgb}{0.9, 0.9, 0.9}

\AtBeginDocument{%
  \providecommand\BibTeX{{%
    \normalfont B\kern-0.5em{\scshape i\kern-0.25em b}\kern-0.8em\TeX}}}

\setcopyright{acmlicensed}
\copyrightyear{2018}
\acmYear{2018}
\acmDOI{XXXXXXX.XXXXXXX}

\acmConference[ASE '24]{Make sure to enter the correct
  conference title from your rights confirmation emai}{October 27 - November 1, 2024}{Sacramento, US}

\renewcommand{\footnotetextcopyrightpermission}[1]{}

\newcommand\nm{LUNAR\xspace}

\newcommand{\ie}{{\em i.e.},\xspace}
\newcommand{\eg}{{\em e.g.},\xspace}

\definecolor{ballblue}{rgb}{0.13, 0.67, 0.8}
\definecolor{jcpink}{RGB}{255, 0, 96}
\collabAuthor{jy}{jcpink}{Jinyang Liu}
\collabAuthor{jc}{jcpink}{JC}
\collabAuthor{jz}{magenta}{Jiazhen Gu}
\collabAuthor{zh}{ballblue}{Zhihan Jiang}
\collabAuthor{jj}{purple}{Junjie}
\collabAuthor{zb}{teal}{Zhuangbin}
\collabAuthor{yc}{blue}{Yichen}

\begin{document}
\title{\nm: Unsupervised LLM-based Log Parsing}

\author{Junjie Huang}
\affiliation{%
  \institution{The Chinese University of Hong Kong}
  \city{Hong Kong}
  \country{China}}

\author{Zhihan Jiang}
\affiliation{%
  \institution{The Chinese University of Hong Kong}
  \city{Hong Kong}
  \country{China}}
\author{Zhuangbin Chen}
\affiliation{%
  \institution{Sun Yat-sen University} \country{China}}
\authornote{Zhuangbin Chen is the corresponding author.}

\author{Michael R. Lyu}
\affiliation{%
  \institution{The Chinese University of Hong Kong}
  \city{Hong Kong}
  \country{China}}

\renewcommand{\shortauthors}{XXX, et al.}

\begin{abstract}

\textit{Log parsing}
serves as an essential prerequisite  for various log analysis tasks.
Recent advancements in this field have improved parsing accuracy by leveraging the semantics in logs through fine-tuning large language models (LLMs) or learning from in-context demonstrations. However, these methods heavily depend on labeled examples to achieve optimal performance. In practice, collecting sufficient labeled data is challenging due to the large scale and continuous evolution of logs, leading to performance degradation of existing log parsers after deployment. To address this issue, we propose \nm, an unsupervised LLM-based method for efficient and off-the-shelf log parsing. Our key insight is that while LLMs may struggle with direct log parsing, their performance can be significantly enhanced through comparative analysis across multiple logs that differ only in their parameter parts. We refer to such groups of logs as \textit{Log Contrastive Units (LCUs)}. Given the vast volume of logs, obtaining LCUs is difficult. Therefore, \nm introduces a hybrid ranking scheme to effectively search for LCUs by jointly considering the \textit{commonality} and \textit{variability} among logs. Additionally, \nm crafts a novel parsing prompt for LLMs to identify contrastive patterns and extract meaningful log structures from LCUs. Experiments on large-scale public datasets demonstrate that \nm significantly outperforms state-of-the-art log parsers in terms of accuracy and efficiency, providing an effective and scalable solution for real-world deployment.\footnote{The code and data are available at \url{https://github.com/Jun-jie-Huang/LUNAR}}.

\end{abstract}

\maketitle

\section{Introduction}

\begin{figure}[t]
    \centering
    \includegraphics[width=0.95\linewidth]{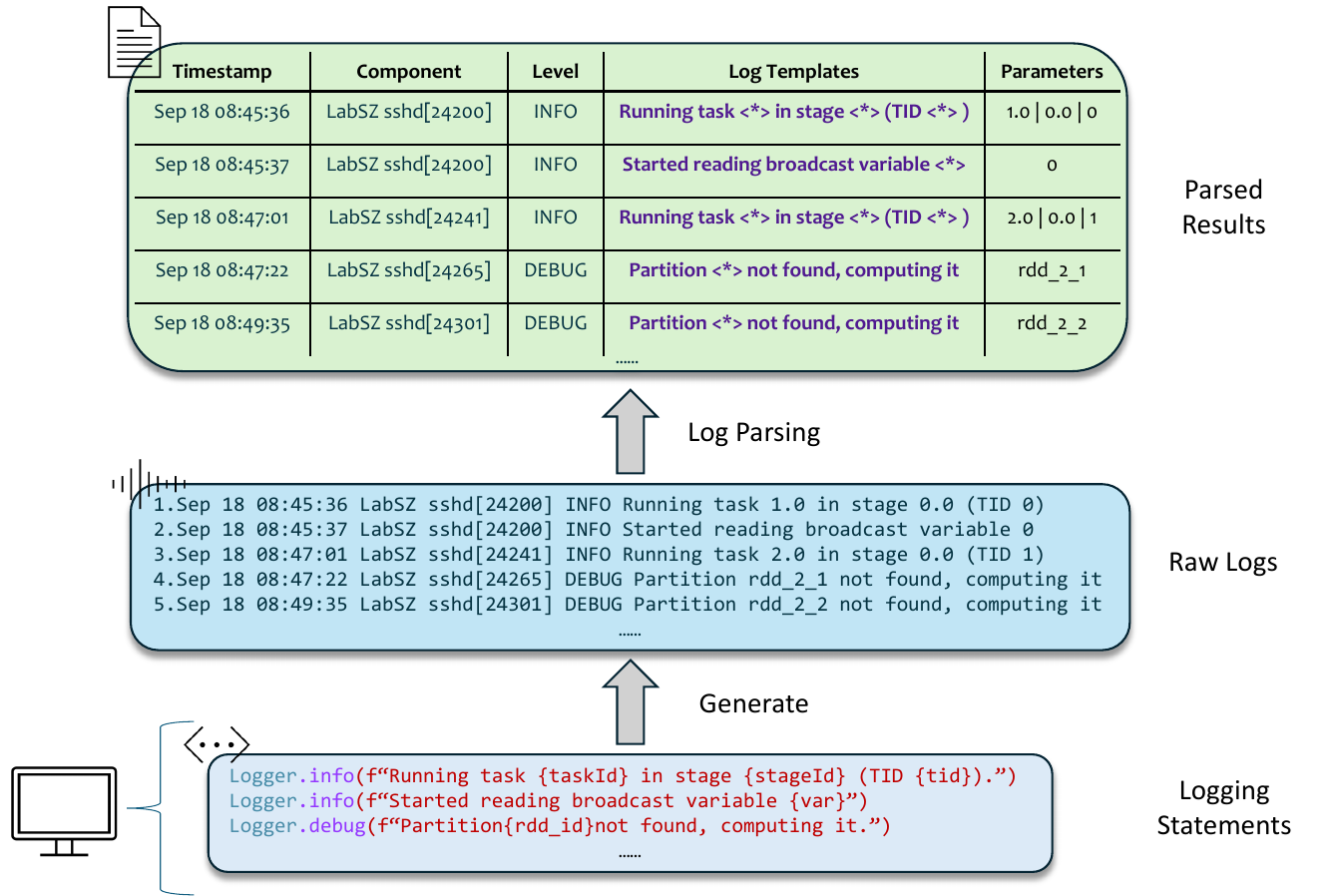}
    \caption{An example of log parsing procedure.}
    \vspace{-12pt}
    \label{fig:parsing_example}
\end{figure}

Log messages record events, transactions, or activities generated by software applications and operating systems at runtime~\cite{he2022empirical,li2023exploring,li2024go}.
They provide valuable insights for system performance  monitoring and reliability assurance.
Various tools have been developed to conduct automated log analysis tasks, including anomaly detection~\cite{zhang2019robust,zhao2021empirical,zhang2022deeptralog,liu2023scalable,chen2021experience}, root cause analysis~\cite{amar2019mining,wang2020root,notaro2023logrule}, and failure diagnosis~\cite{xu2009largescale,chen2021pathidea}.
\textit{Log parsing}, which transforms semi-structured log messages into structured formats~\cite{khan2022guidelines}, serves as a critical preliminary step in log analysis.
Typically, a raw log message contains two parts: 1) \textit{log templates}: constant parts that describe the main content of the logged event; 2) \textit{log parameters}: dynamic parts that contain the parameters (determined at runtime) associated with the event.
Figure~\ref{fig:parsing_example} demonstrates some log messages generated by their logging statements and parsed into structured data.

A straightforward approach to log parsing involves matching raw log messages with their corresponding logging statements in the source code~\cite{shang2012bridging,schipper2019tracing,bushong2020matching}.
However, in practice, source code is not always accessible, particularly for commercial software and third-party libraries.
Consequently, various data-driven log parsers have been proposed to directly extract templates and parameters from raw logs.
They can be generally categorized into two types: \textit{syntax-based parsers} and \textit{semantic-based parsers}.
Syntax-based parsers~\cite{vaarandi2015logcluster,he2017drain,yu2023brain,xu2023hue} resort to statistical or heuristic rules (\eg log length, word frequency) to identify common parts among logs as the templates.
However, these methods often struggle to accurately recognize templates when log messages deviate from the manually-crafted rules.
To address these limitations, deep learning models have been introduced in this field to leverage more advanced features of log data, \ie their semantics.
For example, LogPPT~\cite{le2023log} fine-tunes a pre-trained language model (\eg RoBERTa~\cite{liu2019roberta}) based on manually labelled log templates, aiming for better performance.
With the increasing popularity of large language models (LLMs) for log analysis, some of the latest work~\cite{xu2023prompting,jiang2023llmparser} also employ LLMs for log parsing.
These LLM-based log parsers leverage the extensive pre-trained knowledge of LLMs and utilize the in-context learning (ICL) paradigm~\cite{dong2022survey,gao2023constructing}.
By providing labelled examples as demonstrations, they specialize LLMs for the task of log parsing.

Despite effective, existing semantic-based log parsers largely depend on labelled examples to achieve optimal performance~\cite{ma2024llmparser}.
However, collecting sufficient labelled data is challenging in real world, which hinders their applicability and scalability in practice.
On one hand, production log data are often large in volume, \eg hundreds of millions of logs per hour according to recent studies~\cite{wang2022spine}.
Manually annotating log messages of such scale from diverse systems is labor-intensive and error-prone, which demands substantial domain expertise to ensure accuracy and consistency~\cite{jiang2023large}.
On the other hand, the log data of real-world systems are continuously evolving over time~\cite{wang2022spine,xu2023prompting}.
New log messages and log templates can emerge frequently, reflecting changes in system behavior, updates, and new feature deployments.
This dynamic nature necessitates constant re-annotation and adaptation of the parsers, making it difficult to maintain a high accuracy.
Consequently, the performance of existing semantic-based log parsers may degrade significantly without continuous and significant manual intervention.
As demonstrated in our empirical study (Section~\ref{sec:limitation}), when the proportion of labelled examples decreases by 70\%, the F1 score of template accuracy (FTA) of the state-of-the-art semantic-based log parsers, LogPPT~\cite{le2023log} and LILAC~\cite{jiang2023llmparser}, drops by 33\% and 15\%, respectively.

To address this label-demanding problem, we propose \nm, an \underline{L}LM-based \underline{un}supervised log p\underline{ar}ser, which can generalize to any new log dataset without manual annotations or prior knowledge of the specific log formats.
The core idea behind LUNAR lies in leveraging LLMs' ability to perform comparative analysis on multiple log messages that vary in their parameter parts.
While LLMs may struggle with direct log parsing, the comparison can derive valuable insights by identifying and interpreting patterns across these variations.
For instance, by contrasting logs such as ``\texttt{session opened for user news}'' and ``\texttt{session opened for user test},'' LLMs can easily infer that the tokens ``news'' and ``test'' are likely to represent a username parameter.
We refer to such group of logs as a Log Contrastive Unit (LCU), which are similar enough to facilitate meaningful comparisons, yet diverse enough to highlight the variable parameters.
To find LCUs, we introduce a hybrid ranking scheme by jointly considering the \textit{commonality} and \textit{variability} among logs.
However, given the vast volume of logs, evaluating every possible combination of logs is impractical.
Thus, we introduce a hierarchical sharder to first divide logs into different buckets based on log length and top-$k$ frequent tokens.
The LCU selection and subsequent log parsing can then be efficiently performed in each bucket in parallel.
With the selected LCU, \nm crafts a novel specialized parsing prompt without the need of labelled templates for ICL.
The prompt specifies task intention and output constraints in detail, and provides representative parameter examples to inform LLMs of the parameter characteristics.

We evaluate the performance of \nm against a range of label-free (unsupervised) parsers and label-dependent parsers.
The experiments are conducted on 14 large-scale log parsing datasets in Loghub-2.0~\cite{jiang2023large} from LogPAI~\cite{zhu2019tools}.
The results show that \nm substantially outperforms the unsupervised baselines, surpassing Brain~\cite{yu2023brain} and LILAC w/o ICL~\cite{jiang2023llmparser} by 46.2\% and 26.9\% in FTA, respectively. 
When compared with label-dependent methods, \nm achieves performance on par with the current state-of-the-art parser, LILAC.
Moreover, with parallelization capability, \nm attains a parsing speed comparable to most syntax-based baselines and superior than semantic-based baselines, facilitating efficient parsing of large-scale log data.
Our evaluation shows the potential of \nm for deployment in real-world production systems, where the efficiency, accuracy and generalizability are critical concerns. 

To sum up, the main contributions of this work are as follows:
\begin{itemize}[leftmargin=*]
    \item To the best of our knowledge, we propose the first LLM-based unsupervised log parser dubbed \nm, which instructs LLMs to make comparisons on log contrastive units (LCUs). 
    \item To enable efficient LCU sampling, we introduce a hierarchical sharding scheme, which reduces sample overhead and allows parallel parsing. Moreover, we propose a hybrid method to measure the LCU by balancing the commonality and variability. 
    \item We evaluate \nm on large-scale public datasets. The results show that \nm significantly outperforms unsupervised parsers and achieves comparable performance with state-of-the-art LILAC in terms of accuracy and efficiency. 
\end{itemize}

\section{Motivation}

\subsection{Limitations of Semantic-based Log Parsers}
\label{sec:limitation}
Recently, semantic-based log parsers~\cite{liu2022uniparser, le2023log} have gained significant attention, outperforming traditional syntax-based methods by a considerable margin~\cite{le2023log,jiang2023large}. 
The improvement stems from the use of language models (\eg RoBERTa~\cite{liu2019roberta} and ChatGPT~\cite{openai-api}) to comprehend the semantics of log messages, enabling precise distinction between static template and varying parameters in the log messages. 
However, these parsers require labelled examples (\ie log messages and their corresponding ground-truth log templates) to tailor the language models for log parsing tasks. 
For instance, the state-of-the-art log parser, LILAC~\cite{jiang2023llmparser}, employs labelled samples to perform in-context learning (ICL) to guide language models in producing templates, which is a crucial component to ensure its effectiveness.

Despite the promising results reported, implementing these semantic-based log parsers in practice is challenging due to the difficulty of collecting sufficient labelled data to achieve their full effectiveness.
The reasons behind this are two-fold.
Firstly, manually collecting labelled examples from log data is labor-intensive and error-prone, which requires extensive domain expertise. 
Secondly, software frequently undergoes changes~\cite{zhang2019robust,wang2022spine, xu2023prompting}, resulting in new semantics and patterns in the log data (\ie concept drift~\cite{gama2014survey}). 
Consequently, the performance of these log parsers tends to degrade over time.

\begin{figure}[t]
    \centering
    \includegraphics[width=\columnwidth]{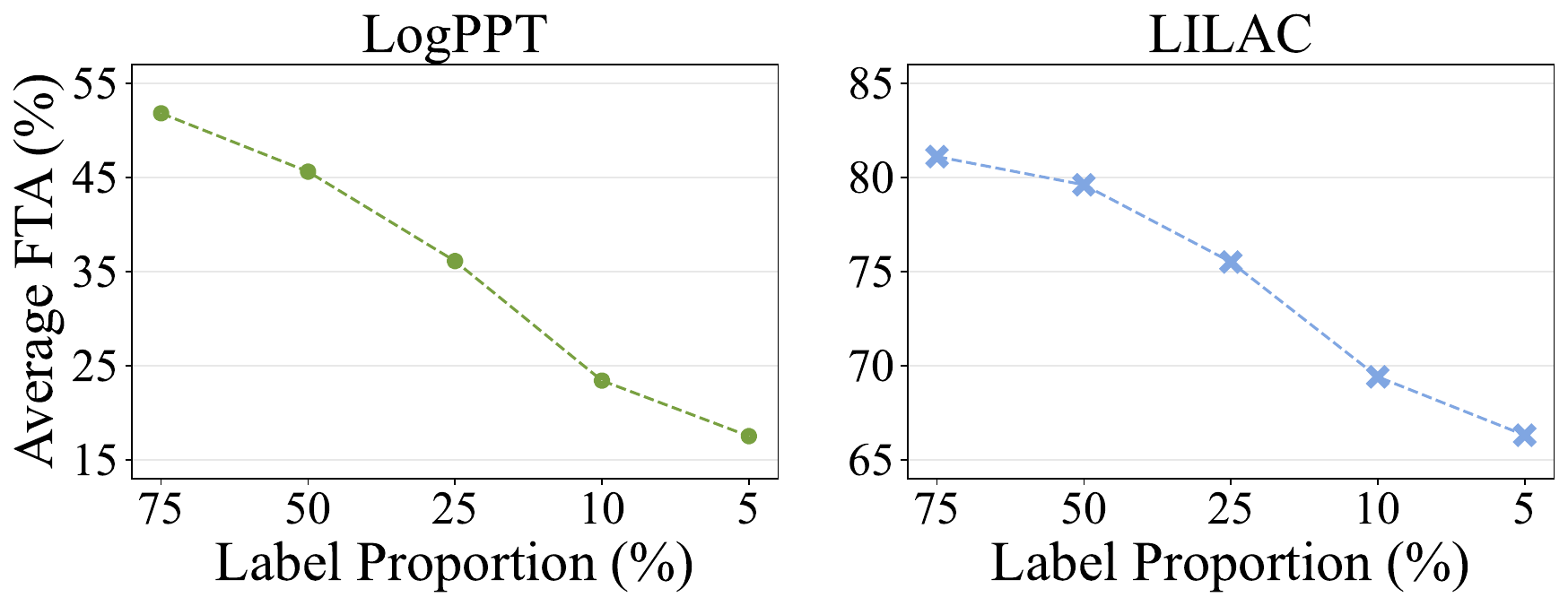}
    \caption{Empirical study on the influences of label proportion on semantic-based log parsers.}
    \label{fig:empirical_study}
\end{figure}

To quantitatively understand the impact of labelled examples on parsing performance, we re-evaluated two representative semantic-based log parsers, \ie LogPPT~\cite{le2023log} and LILAC~\cite{jiang2023llmparser}, using varying label proportions. Specifically, our study utilized large-scale log parsing datasets from Loghub-2.0~\cite{jiang2023large}, which contain 50.4 million log messages from 14 real-world software systems.
For each dataset, we randomly sampled different proportions of labelled log templates as the model-accessible oracles, which are then applied for fine-tuning or ICL. The proportion ranges among 75\%, 50\%, 25\%, 10\%, and 5\%, 
simulating real-world scenarios where oracle labelled templates are scarce due to insufficient manual labeling or system evolution.
We measured their effectiveness using the widely adopted F1 score of template accuracy (FTA)~\cite{khan2022guidelines}. To reduce the bias introduced by randomness, we performed the experiments five times for each setting, following previous studies~\cite{zhu2019tools,jiang2023large,xu2023prompting,le2023log}. We reported the average FTA scores, as shown in Figure~\ref{fig:empirical_study}.

We observe that the performance of both LogPPT and LILAC significantly declines as the label proportion decreases. 
For example, when the label proportion is 75\%, LogPPT achieves a performance of over 50\%.
However, when the label proportion drops to 5\%, its performance falls to about 17\%, indicating a substantial decrease of 33\%.
Similarly, with a label proportion of 75\%, LILAC attains an average FTA score of approximately 81\%. When the proportion decreases to 25\%, the score drops to about 75\%.
At a label proportion of just 5\%, the score further declines to around 66\%, representing a reduction of over 15\% compared to the highest score.
These results suggest that the proportion of labelled examples can significantly impact the performance of semantic-based log parsers, posing challenges to  apply them into real-world software systems.
Therefore, we aim to develop a label-free semantic-based log parser, which can avoid label insufficiency problems and generalize to unseen software without labelled templates.

\begin{figure}[t]
    \centering
    \includegraphics[width=\columnwidth]{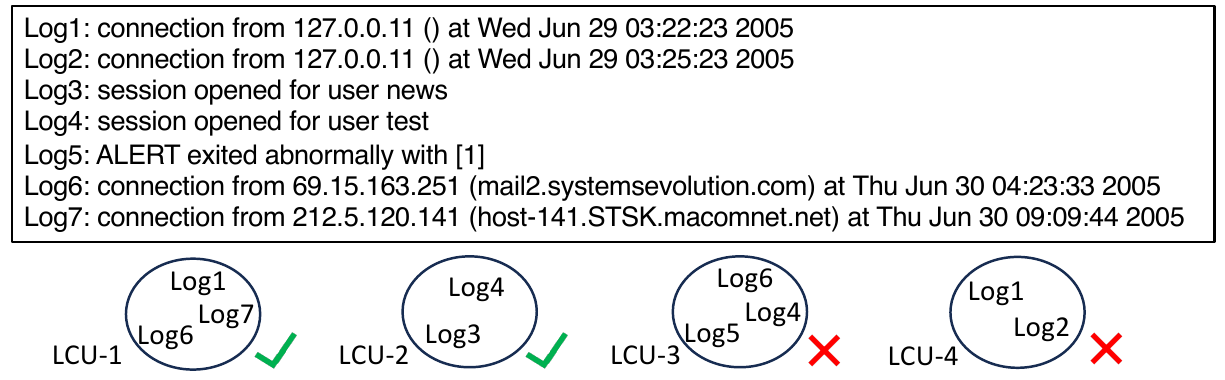}
    \caption{Examples of log contrastive units (LCUs).}
    \label{fig:lcu-example-empirical}
\end{figure}

\subsection{Log Contrastive Unit for LLM-based Parsing}\label{sec:empirical-lcu}
In this section, we present our motivation for developing a label-free, LLM-based unsupervised log parsing approach.

Existing solutions rely on labelled data to guide language models in inferring the parameters from log messages. 
In contrast, \textit{our core insight is that the LLM itself can derive sufficient hints by comparing multiple logs.}
For example, by comparing the logs: ``session opened for user news'' and ``session opened for user test'', LLMs can easily infer that ``news'' and ``test''  are likely username parameters, because these segments vary while the rest of the log message remains consistent. 
We aim to leverage such comparisons to guide LLM without the need for labelled data.

We define such grouped log messages as a \textit{log contrastive unit (LCU)}, which consists of multiple log messages potentially sharing the same template, allowing LLMs to parse them through comparisons. Constructing an effective LCU that guides the LLM to produce accurate parsing results is non-trivial. Specifically, log messages within the same LCU need to exhibit both commonality and variability:

\begin{itemize}[leftmargin=*, topsep=0pt]
    \item Commonality: These log messages should share some common tokens. If they differ significantly, they are likely generated by different logging statements and are not different templates. For example, logs in LCU-3 differ greatly, thus failing to provide proper hints to LLMs.
    \item Variability: These log messages should differ in some of their tokens. If they are all identical, the LLM might mistakenly interpret all tokens as constants. For instance, the IP address in logs of LCU-4 is identical, which can mislead the LLM to regard the parameter parts as constants.
\end{itemize}
LCU-1 and LCU-2 are examples of well-formed LCUs, as log messages within them exhibit both commonality and variability, guiding the LLM to infer the templates and parameters through comparison. However, we observe the following challenges in practice:

\begin{itemize}[leftmargin=*, topsep=0pt]
    \item \textbf{Challenge 1: LCU Volume Explosion}: The sheer volume of log data generated by modern software systems can lead to an exponential increase in the number of potential LCUs. This explosion in combinations makes it challenging to scale the LCU-based approach efficiently. 
    \item \textbf{Challenge 2: Balance of Commonality and Variability}: Log messages often exhibit a high degree of diversity, with templates containing vastly different tokens. This diversity makes it difficult to identify effective LCUs that strike a balance between commonality and variability. Finding the right combinations of log messages that share enough common tokens while still exhibiting variability is a complex task, further complicated by the dynamic nature of log data and the need for continuous adaptation.
\end{itemize}

\section{Methodology}

\begin{figure*}[t]
    \centering
    \includegraphics[width=0.92\linewidth]{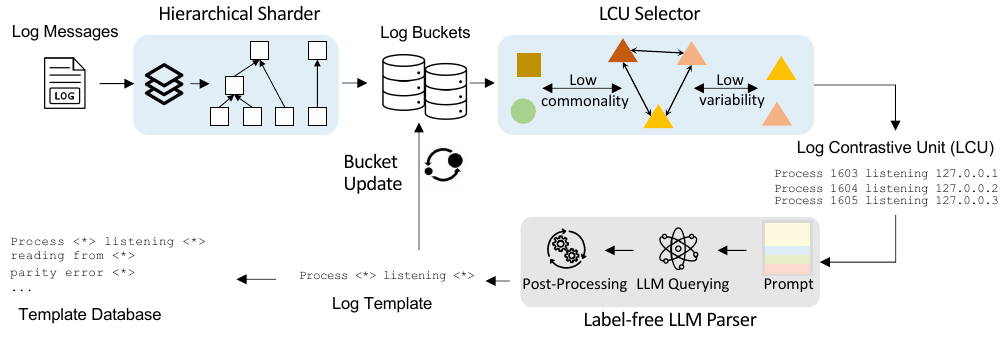}
    \caption{The overall workflow of \nm. }
\label{fig:method-framework}
\end{figure*}

\subsection{Overview}\label{sec:method-overview}
Figure~\ref{fig:method-framework} illustrates the overall framework of \nm, which consists of three main components: \textit{hierarchical sharder}, \textit{generative LCU ranker}, and \textit{label-free LLM parser}.
To address the first challenge, we propose the hierarchical sharder, which divides raw log messages into different buckets based on log length and top-$k$ ranked tokens. By separating logs with low similarity into different buckets, this component reduces the sampling overhead and enables parallelization for efficient parsing (\S\ref{sec:method-sharding}).
To address the second challenge, we propose the generative LCU ranker, which operates within each bucket to continuously sample LCUs for the LLM to parse. This module computes a hybrid LCU score that jointly considers both the commonality and variability of the LCUs, guiding the sampling process to ensure effective LCUs (\S\ref{sec:method-sampling}).
Lastly, the label-free LLM parser constructs an organized prompt for the mined LCUs to query the LLM and obtain the templates from the LLM's response. We introduce a novel format for organizing the parsing prompt, which specifies the task intention and output constraints, and includes several representative parameter examples to inform the LLM of the parameter characteristics (\S\ref{sec:method-llm}).
The combination of all components ensures that the LLM can effectively parse the logs without the need for labelled data, addressing both scalability and diversity challenges.

\begin{figure}[t]
    \centering
    \includegraphics[width=0.98\columnwidth]{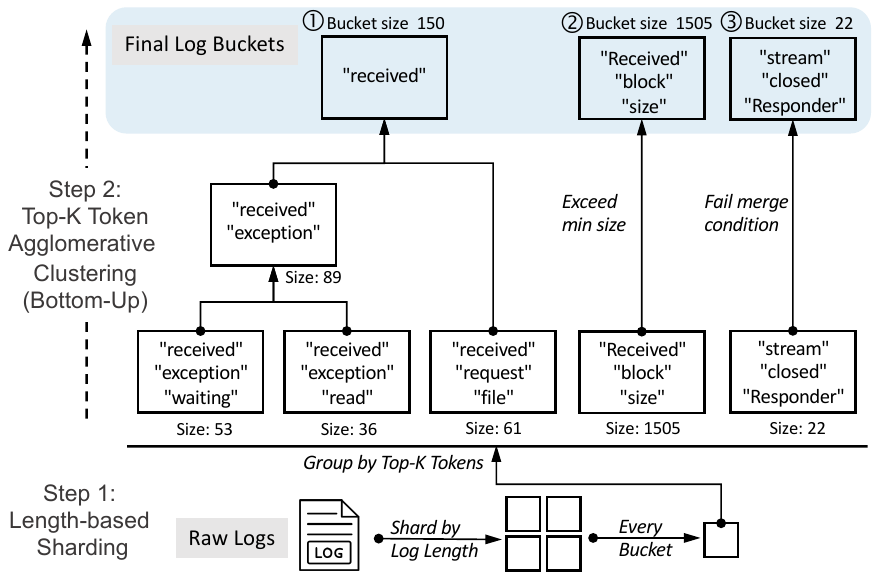}
    \caption{Hierarchical log sharding in \nm. }
    \label{fig:sharding}
\end{figure}

\subsection{Hierarchical Log Sharder}\label{sec:method-sharding}
Directly sampling LCUs from the whole log messages is overwhelming due to the large volume of log data, which often contain million of log messages.
Hence, \nm first groups the log messages into multiple log buckets, where each bucket contain logs that share some similarity.
By doing so, extremely different logs, which have less potentials to belong to the same template~\cite{he2017drain}, are placed to different buckets.
Specifically, we propose a hierarchical algorithm to shard log: 1) firstly divide the log messages by log length, and 2) progressively divide logs by top-$k$ frequent tokens.
Figure.~\ref{fig:sharding} demonstrates the workflow of hierarchical log sharder.

\subsubsection{Length-based Sharding.}
\nm first shards the log messages by the log length.
Logs with the same number of tokens are grouped in a bucket, which is a widely adopted heuristic to conduct initial grouping in log analysis~\cite{he2017drain, li2023logshrink}.
Here we simply adopt the white-space as the delimiter.
We do not use other delimiters such as `:` because most of these delimiters appear in parameters and introducing them can produce over-fragmented logs, leading to a noisy sharding result.

\subsubsection{Top-$k$ Token Agglomerative Clustering.}
However, simply sharding by log length is too coarse-grained since logs belonging to different templates could have the same length~\cite{jiang2023llmparser}. 
Thus \nm continuously shards each bucket into more purified ones where less templates are included in one bucket. 
Intuitively, the static parts of a log generally have higher occurrences than its parameter part, therefore logs that share the most frequent tokens have more potential to belong to the same template~\cite{liu2019logzipIC, jiang2023large, jiang2023llmparser}. 
To achieve this, we introduce a \textit{hierarchical agglomerative clustering} algorithm based on top-$k$ frequent tokens, which builds the sub-buckets in a bottom-up way. 

Specifically, for each bucket obtained from the first step, we first group logs into singleton clusters based on the same top-$k$ tokens in each log.
The top-$k$ tokens of a log is an ordered list extracted based on the token frequency and position.
The frequency is counted among logs in the bucket.
The more frequent tokens takes a more preceding place in the list.
However, some log messages can have multiple tokens with the same frequency, which leads to confusion in ranking; hence we additionally rank tokens with the same frequency by their positions in the logs, with the preceding token being ranked before the following token.

Then, we iteratively merge the singleton clusters to obtain final log buckets until a stopping criterion is met. 
In each iteration, if a cluster contains more than $N$ logs, it will be directly nominated as a standalone bucket and no further merge operations will be conducted on it. 
For the remaining clusters, which contain less than $N$ logs, we merge those with the same top-($k-1$) tokens to form a parent cluster. 
The iteration stops if all clusters exceed $N$ logs, or no any two clusters satisfy the merge condition. 
Finally, we obtain a collection of buckets containing a smaller number of logs with higher relevance. 

It is worth noting that the collected log buckets have a two-level hierarchy, where each hyper-bucket obtained by length-based sharding is disjointly separated to multiple final buckets. Therefore, buckets under the same level are mutually exclusive with each other, \ie every log belongs to one bucket on either sharding level. 
The advantages of this design are two-fold. Firstly, once a template is obtained, it can be leveraged to match the logs in the bucket while not required for other buckets.  
Secondly, this characteristic enables parallel parsing, where the buckets can be independently allocated on multiple executors. More details will be described in Section~\ref{sec:method-parallel}

\subsection{Two-stage LCU Selector}\label{sec:method-sampling}
After log sharding, \nm iteratively select a group of similar logs for the LLM to make comparisons and extract templates.
A suitable group of logs should be similar in tokens to each other, as logs generated by the same logging statement share several identical tokens.
On this basis, the logs are expected to have as much variance as possible, enabling LLMs to make cross-log comparisons to infer parameters.
For example, in Figure~\ref{fig:lcu-example}, group \ding{192} and \ding{193} are preferred than group \ding{194} as their logs are more likely to belong to the same template.
Moreover, group \ding{192} is better than \ding{193} due to its higher variance reflected in the diverse parameter value of process index.

We term such a group of logs as a log contrastive unit (LCU).
Logs in each LCU have both commonality (\eg with common words or belonging to the same template) as well as variability (\eg with different tokens), so that it can be leveraged to infer the parameters. 
To collect LCUs, we leverage a two-step sampling approach: \textit{stratified LCU generation} and \textit{hybrid LCU nomination}.

\subsubsection{Stratified LCU Generation.} 
Given a bucket of logs, \nm first generate multiple candidate LCUs that share some similarity.
A straightforward approach is to enumerate all possible log combinations.  However, this method is computationally expensive due to the sheer volume of logs and the complexity of enumeration. To reduce computation overhead, we apply stratified sampling to collect a limited-sized log pool by sampling from different similarity levels before perform combination. 

Given a log bucket, \nm first randomly selects a log message in the bucket as the anchor log.
Next, \nm computes the pair-wise similarities of the anchor log to the remaining logs.
We use Jaccard similarity, which is a widely adopted metric to measure pair-wise log similarity~\cite{he2017drain,jiang2023llmparser}.
Specifically, we first split the log~$l$ into tokens $T(l)$ with white-space delimiter and then compute the Jaccard similarity $JS(l_1, l_2)=\frac{T(l_1) \cap T(l_2)}{T(l_1) \cup T(l_2)}$. 
Then, we remove the logs with a similarity below a threshold $s$ or equals to $1.0$. 
The poor similarity indicates the less possibility to belong to the same template, while a similarity of $1.0$ indicates duplication. 
After that, we create a pool of logs by stratified sampling from each similarity level to ensure balanced sampling.
Specifically, we set the sample size of each level equal to the LCU size $m$ to involve more diverse logs. 
Lastly, from the log pool, \nm generates multiple LCUs by combinatorial enumeration. 
Each LCU consists of an anchor log, as well as $m-1$ logs chosen from the log pool. 
Since we sample a small size of LCU (\eg $m=3$ or $m=4$) from a limited size of log pool, the computational cost of combinatorial enumeration is within an acceptable range (More details about efficiency can found in  Section~\ref{sec:experiment-rq3}).
If the log pool contain less than $m-1$ logs, we directly return the anchor with the pool as the candidate LCU. 
Finally, a candidate set of LCUs are generated and proceeded to hybrid rankding in the next step.

\subsubsection{Hybrid LCU Ranking}
After obtaining a small number candidate LCUs, \nm conducts hybrid ranking to select the most suitable LCU. The selected logs are required to exhibit variability, while maintain some commonality as we discussed in Section~\ref{sec:empirical-lcu}. By doing so, LLMs can be less likely to incorrectly recognize the frequent tokens as parameters, while also benefit from the contrastive information in the LCU. Specifically, we propose to compute the variability score and the commonality score to measure an LCU. 

Firstly, the variability score of an LCU is computed as average of pair-wise distances to measure the overall variability of the logs. Suppose an LCU consists of $L$ logs $LCU=\{l_1, l_2, ... l_L\}$, we first compute the distance between any two logs in the LCU based on Jaccard similarity:
\begin{equation}
    dist(l_i, l_j) = 1 - JS(l_i, l_j).
\end{equation}
Then we average the distances of every two logs in the LCU to obtain the variability score:
\begin{equation}
    S_{LCU}^{Variability} = \frac{2}{L(L-1)}\sum_{i=1}^{L}\sum_{j=i+1}^{L} dist(l_i, l_j), 
\end{equation}
where the higher the variability score, the more diverse the LCU, indicating more contrastive information is included.

\begin{figure}[t]
    \centering
    \includegraphics[width=1.0\columnwidth]{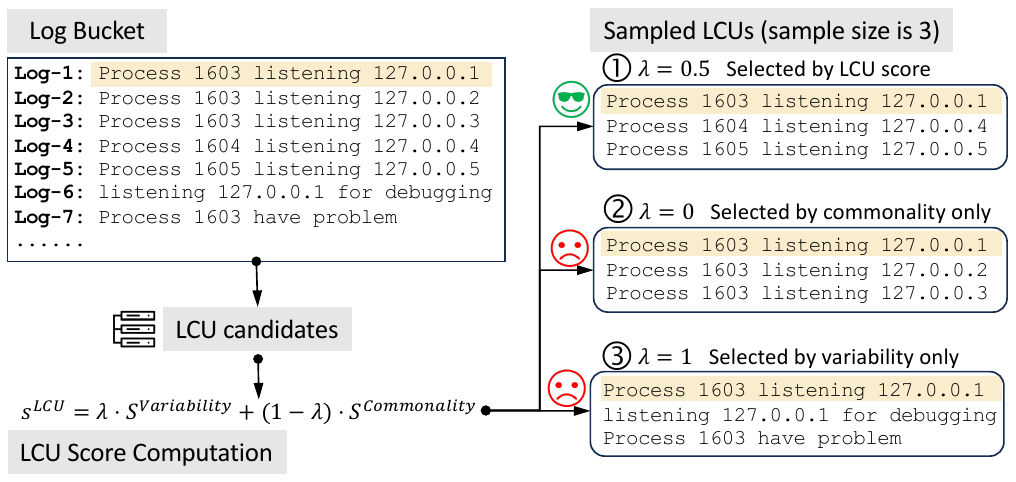}
    \caption{An example of expected LCU in LCU nomination.}
    \label{fig:lcu-example}
\end{figure}

Simply ranking by variability could result in LCUs containing totally different logs (\eg LCU \ding{194} in Figure~\ref{fig:lcu-example}).
As the compensation, the commonality score of an LCU is introduced, which is computed as the average of absolute similarity difference to balance the variance of logs. Specifically, we $P$ log pairs can be obtained from an LCU, then:
\begin{equation}
    S_{LCU}^{Commonality} = \frac{2}{P(P-1)}\sum_{i=1}^{P}\sum_{j=i+1}^{P} (1 - |JS(p_i) - JS(p_j)|),
\end{equation}
\noindent where $JS(p_i)$ is the Jaccard similarity of log pair $p_i$.
The higher the commonality score, the more likely that logs belong to the same template, as the similarity difference between pair to pair is relatively low.
Logs that share the same template but differ only in their parameters will exhibit exactly such a pattern.
For example, in the LCU \ding{193} of Figure~\ref{fig:lcu-example}, the similarity between any two logs is the same as any other two logs within the LCU, which implies the highest commonality score of 1.

To jointly consider the variability and commonality scores, we combine them with linear interpolation with a weight of $\lambda$, to obtain the hybrid LCU score.
The LCU with maximum interpolation score selected as the final LCU, which is computed as follows:

\begin{equation}
    S_{LCU} = \lambda \cdot S_{LCU}^{Variability} + (1-\lambda) \cdot S_{LCU}^{Commonality}.
\end{equation}

To present the LCU score in a straightforward way, we provide an example with a sample size of $3$ in  Figure~\ref{fig:lcu-example}. 
In this example, by selecting with variability score only, LCU \ding{194} will be sampled. However, LCU \ding{194} is sub-optimal as the logs belong to three different templates. LLMs might mis-classify the parameter~``1603'' as this group fails to provide useful contrasts towards it.
With the commonality score only, LCU \ding{193} will be sampled. However, this is also sub-optimal due to the identical parameter~``1603''. This LCU could mislead LLMs to incorrectly recognize parameters due the identical tokens.
In contrast, by combining the variability score and commonality score, we can obtain an optimal LCU, where each parameter has at least two values present. Using this LCU can be maximumly inform LLM to consider the changing tokens, thereby leading to an improved parsing accuracy.

\subsection{Label-free LLM Parser}\label{sec:method-llm}
Upon obtaining an LCU, \nm applies an LLM-based parser to extract the template without any labelled examples. 
The parser leverages the strong textual comprehension ability of LLMs, which have shown remarkable few-shot and zero-shot performance in various information extraction tasks, such as named entity recognition~\cite{xie2023zeroner} and text classification~\cite{lyu2023zicl}. Therefore, we believe LLMs have the potential to solve log parsing in an unsupervised way. 
Specifically, our LLM parser first creates a well-designed prompt for each LCU to instruct the LLM~(\S\ref{sec:method-llm-prompt}) and then extract the template from the LLM response~(\S\ref{sec:method-llm-template}).

\subsubsection{Prompt Design}\label{sec:method-llm-prompt}
\begin{figure}[t]
    \centering
    \includegraphics[width=0.9\columnwidth]{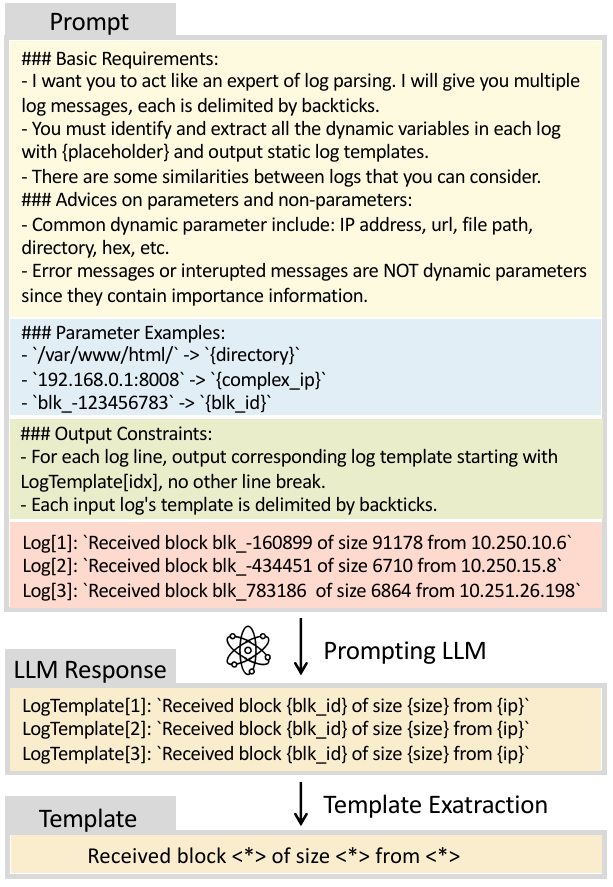}
    \caption{Demonstration-free parsing prompt in \nm.}
    \label{fig:prompt}
    \vspace{-10pt}
\end{figure}

As LLMs are not specifically tuned for log parsing, existing LLM-based log parsers leverage exemplars of log and its labelled template to instruct the task~\cite{jiang2023llmparser, xu2023prompting}. These demonstrations could specify task intents, output format, and parameter information of the domain, which are useful to enhance log parsing~\cite{ma2024llmparser, min2022rethinking}. 
However, in this work, we focus on developing an unsupervised parser where no labelled templates are provided. Therefore, we need to design a more concrete prompt covering the aforementioned demanding information to facilitate accurate parsing.
Specifically, our prompt contains four parts: task instructions, parameter examples, output constraints, and input LCUs. Figure~\ref{fig:prompt} shows a prompt example.

\noindent \textbf{Task Instruction.}
A clear and useful prompt should provide comprehensive instructions for LLMs to understand the task~\cite{he2022rethinking}. 
In LUNAR, we construct the task instruction of log parsing with two parts: basic requirements and general advice on parameters. 
For the basic requirements, the task input/output and the mapping mechanism from input to output are firstly specified. The instruction also guides the LLM to consider the similarity between the  grouped input logs. 
For the general advice on parameters, high-level descriptions of parameter categories are provided. We also instruct the LLM not to consider the error messages as parameters since they also contain some important information.

\noindent \textbf{Parameter Examples.}
Inspired by the entity examples used for unsupervised NER~\cite{jiang2024picl}, we involve parameter examples to the prompt. 
These examples are used to provide domain knowledge and instruct the task, which consist of a parameter type and a parameter value. For instance, ``\texttt{/var/www/html/}'' is the value of a parameter ``directory.'' 
To obtain the parameter examples, we first manually check the regular expressions of used in previous studies~\cite{he2017drain, li2023did}, then sample a value from matched parameters, and finally summarize the parameter type of the value~\cite{li2023did}. 
In the prompt, the parameter value and type are connected by an arrow, aiming to indicate the task intent to convert the value to a placeholder with the bracket.

\noindent \textbf{Output Constraints.}
In output constraints, we explain the desired output formats. The prompt instructs the LLM to produce a template for every log message. Each generated template should be appended by a prefix of ``LogTemplate[idx]'' and delimited with backticks to facilitate template aggregation and extraction.

\noindent \textbf{Queried LCUs.} 
Finally, the LCU with multiple logs are incorporated to the prompt. We arrange the logs in a sequential order and apply a prefix schema (\ie ``Log[]:'') and index indicator to indicate the order. In this way, the LLM can make straightforward comparisons and produce corresponding template for each log.

Guided by the task instruction, parameter examples, and output constraints, the LLM could more accurately generate the templates of the queried logs in the LCU. 

\subsubsection{Template Acquisition}\label{sec:method-llm-template}
After constructing the prompt, we leverage the LLM to identify the templates in the LCU. 
Specifically, we first leverage prompt to query an LLM to get a response. 
After that, we perform post-processing to extract valid templates from the response. 
Specifically, we extracts all backticked strings with a ``LogTemplate'' prefix, thanks to clear output constraints in the prompt. Then we aggregate the extracted templates into one template by selecting the most frequent template. Finally, we replace the bracketed parameters by the placeholder $<*>$.

\subsection{Bucket Updating}\label{sec:method-update}
After obtaining a valid template, \nm examines the logs that match the template. The template is assigned as the final prediction to these successfully matched logs, which are then removed from the buckets. The remaining log buckets are used for next iteration of parsing. Once all buckets are empty, the parsing terminates. 

It is worth-noting that the the hierarchical sharding produces buckets at two levels, \ie sharded by length and further clustered by top-$k$ tokens. A straightforward approach for bucket updating is to only examine and update the final bucket. However, logs in the same template can be placed in different buckets of equal length, potentially causing redundant iterations for one template,  leading to prolonged parsing time and increased LLM querying expenses.
To balance the parsing overhead, \nm conducts template examination and updating on the buckets under the first sharding level, meaning that buckets with identical log lengths are all examined once a template is obtained.

\subsection{Parallelization}\label{sec:method-parallel}
In previous sections, we have discussed the building components of \nm. In practice, scalability is a crucial issue when parsing vast amounts of log messages~\cite{wang2022spine}. Therefore, we introduce the parallelization mechanism of \nm to improve the parsing speed. 

As mentioned in \S\ref{sec:method-sharding}, hierarchical sharding produces mutually exclusive log buckets at two levels. Consequently, logs in different buckets can be processed in a completely parallel fashion. However, since bucket updating is performed among all the buckets of the same length at the first level. Simply allocating buckets at the second level to produce template and update buckets can lead to \textit{template conflict}, \ie different templates being assigned to the same log. 
To address the problem, the allocation is conducted at the first level. 
Specifically, we aim to allocate $n$ executors for all first-level log buckets. Within each first-level log bucket, \nm iteratively selects the largest second-level bucket, selects an LCU, queries the LLM to obtain the template, and updates buckets under this level.

\section{Experiment Setup}
We conduct extensive experiments to evaluate \nm by answering the following research questions (RQs):
\begin{itemize}[leftmargin=*, topsep=0pt]
    \item \textbf{RQ1:} How effective is \nm?
    \item \textbf{RQ2:} How do different settings affect \nm?
    \item \textbf{RQ3:} How efficient is \nm?
\end{itemize}

\subsection{Datasets}
We evaluate \nm on Loghub-2.0~\cite{he2020loghub,jiang2023large}, a collection of large-scale benchmark datasets for log parsing provided by LogPAI team~\cite{zhu2019tools}. 
Loghub-2.0 is annotated with ground-truth templates for 14 diverse log datasets, which covers a variety of systems such as distributed systems, supercomputer systems, and server-side applications.
Compared with the original Loghub~\cite{he2020loghub}, Loghub-2.0 is equipped with $3\times$ templates and $1,800\times$ log messages on average, which can support more comprehensive evaluation on accuracy and efficiency.  
On average, each dataset in Loghub-2.0 comprises 3.6 million log messages. 
In total, the collection contains approximately 3,500 unique log templates. 

\subsection{Baselines}
We compare \nm with seven open-sourced state-of-the-art log parsers, including four unsupervised syntax-based parsers and three label-required semantic-based parsers.
For unsupervised label-free parsers, we adopt three syntax-based methods and one LLM-based method. The syntax-based methods include AEL~\cite{jiang2008abstracting}, Drain~\cite{he2017drain}, and Brain~\cite{yu2023brain}, chosen for their superior performance compared to other syntax-based methods~\cite{zhu2019tools,khan2022guidelines,jiang2023large}.
As there are currently no LLM-based log parsers proposed for label-free parsing, we adopt a label-free variant of LILAC~\cite{jiang2023llmparser} (LILAC w/o ICL) by removing the in-context learning (ICL) module.
In terms of label-required log parsers, we select three state-of-the-art semantic-based log parsers: UniParser~\cite{liu2022uniparser}, LogPPT~\cite{le2023log}, and LILAC~\cite{jiang2023llmparser}.
UniParser trains a long short-term memory (LSTM) model on labelled log data for log parsing. 
LogPPT uses labelled log data to perform prompt-based fine-tuning based on RoBERTa~\cite{liu2019roberta}.
LILAC samples similar logs from a small set of labelled logs (\eg 32 logs) for each log, utilizing an ICL paradigm for LLMs to parse logs.

\subsection{Evaluation Metrics}

Following previous works~\cite{zhu2019tools,khan2022guidelines,jiang2023large},  we evaluate \nm with the following four metrics. 
\begin{itemize}[leftmargin=*]
  \item \textit{Grouping Accuracy} (GA) ~\cite{zhu2019tools} is a log-level metric that measures the the amount of log messages of a same template are grouped together by the parser. It is computed as the ratio of correctly grouped log messages over all log messages, where a log message is regarded as correctly grouped if and only if its predicted template have the same group of log messages as the oracle. 
  \item Parsing Accuracy (PA)~\cite{dai2020logram} is a log-level metric that measures the correctness of extracted templates and variables. It is defined as the ratio of correctly parsed log messages over all log messages, where a log message is considered to be correctly parsed if and only if all its static text and dynamic variables are identical with the oracle.
  \item F1 score of Grouping Accuracy (FGA)~\cite{jiang2023large} is a template-level metric that measures the ratio of correctly grouped templates. It is computed as the harmonic mean of precision and recall of grouping accuracy, where the template is considered as correct if log messages of the predicted template have the same group of log messages as the oracle. 
  \item F1 score of Template Accuracy (FTA)~\cite{khan2022guidelines} is a template-level accuracy computed as the harmonic mean of precision and recall of Template Accuracy. A template is regarded as correct if and only if it satisfies two requirements: log messages of the predicted template have the same group of log messages as the oracle, and all tokens of the template is the same as the oracle template.
\end{itemize}

\subsection{Environment and Implementation}
We conduct all the experiments on a Ubuntu 20.04.4 LTS server with 256RAM and an NVIDIA A100 40G GPU. 
The LLM used in \nm is ChatGPT (gpt-3.5-turbo-0613) due to its popularity in recent log analysis studies~\cite{xu2023prompting, jiang2023llmparser, ma2024llmparser}. We invoke the official API provided by OpenAI~\cite{openai-api} and set the temperature to 0 to avoid randomness in token generation and ensure reproducibility. 
By default, we set the a top-$k$ token number to 3 and minimum cluster size to 100 in our hierarchical log sharder. For the LCU selector, we use an LCU sample size of 3, a interpolation factor $\lambda$ of 0.7, and a minimum similarity of 0.33, respectively.
For parallelization, we distribute the jobs to 8 executors. 
We also conduct experiments of \nm with different LCU sample sizes and interpolation factors.
As for baseline methods, we directly use the accuracy reported in previous work~\cite{jiang2023large} and rerun them with their default parameters on the same environment to fairly compare the efficiency.

\section{Evaluation Results}

\begin{table*}[]
\captionsetup{justification=centering}
\centering
\caption{Accuracy of \nm compared to state-of-the-art baselines on public datasets. (\%)}
\label{tab:RQ1}
\resizebox{\textwidth}{!}{%
\begin{NiceTabular}{c||c|cccccccccccccc|c}
\toprule
{Method} & {Metric} & {\small Proxifier} & {\small Apache} & {\small OpenSSH} & {\small HDFS} & {\small OpenStack} & {\small HPC} & {\small Zookeeper} & {\small HealthApp} & {\small Hadoop} & {\small Spark} & {\small BGL} & {\small Linux} & {\small Mac} & {\small Thunderbird} & {\small Average} \\

\midrule
\rowcolor{grey}\multicolumn{17}{c}{\textbf{Unsupervised Log Parsers}} \\
\midrule

\multirow{4}{*}{AEL} & GA & 97.4 & \textbf{100.0} & 70.5 & \underline{99.9} & 74.3 & 74.8 & 99.6 & 72.5 & 82.3 & — & 91.5 & 91.6 & 79.7 & 78.6 & 85.6 \\
                     & PA & 67.7 & 72.7 & 36.4 & 62.1 & 2.9 & 74.1 & 84.2 & 31.1 & 53.5 & — & 40.6 & 8.2 & 24.5 & 16.3 & 44.2 \\
                     & FGA & 66.7 & \textbf{100.0} & 68.9 & 76.4 & 68.2 & 20.1 & 78.8 & 0.8 & 11.7 & — & 58.7 & 80.6 & 79.3 & 11.6 & 55.5 \\
                     & FTA & 41.7 & 51.7 & 33.3 & 56.2 & 16.5 & 13.6 & 46.5 & 0.3 & 5.8 & — & 16.5 & 21.7 & 20.5 & 3.5 & 25.2 \\
\hline
\multirow{4}{*}{Drain} & GA & 69.2 & \textbf{100.0} & 70.7 & \underline{99.9} & \underline{75.2} & 79.3 & 99.4 & 86.2 & \underline{92.1} & 88.8 & 91.9 & 68.6 & 76.1 & \underline{83.1} & 84.3 \\
                       & PA & 68.8 & 72.7 & 58.6 & 62.1 & 2.9 & 72.1 & 84.3 & 31.2 & 54.1 & 39.4 & 40.7 & 11.1 & 35.7 & 21.6 & 46.8 \\
                       & FGA & 20.6 & \textbf{100.0} & \underline{87.2} & \underline{93.5} & 0.7 & 30.9 & 90.4 & 1.0 & 78.5 & 86.1 & 62.4 & 77.8 & 22.9 & 23.7 & 55.4 \\
                       & FTA & 17.6 & 51.7 & 48.7 & 60.9 & 0.2 & 15.2 & 61.4 & 0.4 & 38.4 & 41.2 & 19.3 & 25.9 & 6.9 & 7.1 & 28.2 \\
\hline
\multirow{4}{*}{Brain} & GA & 52.1 & \underline{99.7} & 66.3 & 96.0 & \textbf{100.0} & 80.0 & 99.3 & 97.9 & 56.3 & 97.2 & \underline{94.0} & 79.0 & \underline{83.4} & 79.2 & 84.3 \\
                       & PA & \underline{70.3} & 28.7 & 48.1 & 92.9 & 14.1 & 66.3 & 82.2 & 17.5 & 14.3 & 39.3 & 40.2 & 1.0 & 32.5 & 26.1 & 41.6 \\
                       & FGA & 73.7 & \underline{93.3} & 75.9 & 75.9 & \textbf{100.0} & 44.7 & 79.8 & 87.2 & 52.8 & 20.8 & 75.6 & 75.1 & 75.4 & 74.8 & 71.8 \\
                       & FTA & 73.7 & 46.7 & 34.5 & \underline{62.1} & 29.2 & 21.3 & 60.1 & 33.9 & 20.0 & 1.0 & 19.7 & 27.5 & 29.4 & 27.4 & 35.4 \\
\hline
\multirow{4}{*}{LILAC w/o ICL} & GA & 0.0 & \underline{99.7} & \underline{74.4} & \textbf{100.0} & \textbf{100.0} & 85.6 & \underline{99.7} & 99.3 & 91.5 & \underline{99.8} & 88.5 & 80.4 & 74.2 & 72.9 & 83.3 \\
                               & PA & 11.4 & 94.2 & 34.8 & 94.7 & 47.7 & 64.8 & 35.1 & 53.6 & 71.8 & 58.1 & 82.3 & 67.6 & 48.9 & 53.9 & 58.5 \\
                               & FGA & 0.0 & \underline{93.3} & 69.7 & 65.8 & \textbf{100.0} & \underline{89.0} & \underline{96.0} & \underline{96.2} & \underline{91.5} & 89.0 & 83.3 & 82.5 & 77.8 & 32.8 & 76.2 \\
                               & FTA & 16.0 & \underline{56.7} & 42.4 & 52.1 & 79.2 & 74.0 & 73.1 & 68.2 & 64.5 & 58.9 & 63.7 & 56.5 & 41.6 & 18.4 & 54.7 \\

\midrule
\rowcolor{grey}\multicolumn{17}{c}{\textbf{Label-required Log Parsers}} \\
\midrule

\multirow{4}{*}{UniParser} & GA & 50.9 & 94.8 & 27.5 & \textbf{100.0} & \textbf{100.0} & 77.7 & 98.8 & 46.1 & 69.1 & 85.4 & 91.8 & 28.5 & 73.7 & 57.9 & 71.6 \\
                           & PA & 63.4 & 94.2 & 28.9 & 94.8 & 51.6 & 94.1 & \textbf{98.8} & 81.7 & \textbf{88.9} & 79.5 & 94.9 & 16.4 & \textbf{68.8} & \textbf{65.4} & 73.0 \\
                           & FGA & 28.6 & 68.7 & 0.9 & \textbf{96.8} & \underline{96.9} & 66.0 & 66.1 & 74.5 & 62.8 & 2.0 & 62.4 & 45.1 & 69.9 & 68.2 & 57.8 \\
                           & FTA & 45.7 & 26.9 & 0.5 & 58.1 & 28.9 & 35.1 & 51.0 & 46.2 & 47.6 & 1.2 & 21.9 & 23.2 & 28.3 & 29.0 & 31.7 \\
\hline
\multirow{4}{*}{LogPPT} & GA & \underline{98.9} & 78.6 & 27.7 & 72.1 & 53.4 & 78.2 & 96.7 & \underline{99.8} & 48.3 & 47.6 & 24.5 & 20.5 & 54.4 & 56.4 & 61.2 \\
                        & PA & \textbf{100.0} & 94.8 & 65.4 & 94.3 & 40.6 & \textbf{99.7} & 84.5 & \textbf{99.7} & 66.6 & 95.2 & 93.8 & 16.8 & 39.0 & 40.1 & 73.6 \\
                        & FGA & \underline{87.0} & 60.5 & 8.1 & 39.1 & 87.4 & 78.0 & 91.8 & 94.7 & 52.6 & 37.4 & 25.3 & 71.2 & 49.3 & 21.6 & 57.4 \\
                        & FTA & \underline{95.7} & 36.8 & 10.5 & 31.2 & 73.8 & 76.8 & 80.9 & 82.2 & 43.4 & 29.9 & 26.1 & 42.8 & 27.4 & 11.7 & 47.8 \\
\hline
\multirow{4}{*}{LILAC} & GA & \textbf{100.0} & \textbf{100.0} & 69.0 & \textbf{100.0} & \textbf{100.0} & \textbf{86.9} & \textbf{100.0} & \textbf{100.0} & 87.2 & \textbf{100.0} & 89.4 & \textbf{97.1} & \textbf{87.6} & 80.6 & \underline{92.7} \\
                       & PA & \textbf{100.0} & \underline{99.6} & \textbf{94.1} & \underline{99.9} & \textbf{100.0} & 70.5 & 68.7 & 72.9 & 83.2 & \underline{97.3} & \underline{95.8} & \underline{76.5} & \underline{63.8} & 55.9 & \underline{84.2} \\
                       & FGA & \textbf{100.0} & \textbf{100.0} & 83.8 & \textbf{96.8} & \textbf{100.0} & \textbf{90.7} & \textbf{96.7} & \textbf{98.1} & \textbf{96.2} & \underline{90.1} & \underline{85.9} & \textbf{93.1} & \textbf{82.5} & \underline{79.3} & \textbf{92.4} \\
                       & FTA & \textbf{100.0} & \textbf{86.2} & \underline{86.5} & \textbf{94.6} & \textbf{97.9} & \underline{80.0} & \underline{86.8} & \underline{87.3} & \textbf{77.9} & \underline{75.9} & \underline{74.6} & \textbf{74.0} & \underline{55.3} & \underline{57.2} & \underline{81.0} \\

\midrule
\rowcolor{grey}\multicolumn{17}{c}{\textbf{Our proposed method }} \\
\midrule

\multirow{4}{*}{LUNAR} & GA & \textbf{100.0} & \textbf{100.0} & \textbf{78.0} & \textbf{100.0} & \textbf{100.0} & \underline{86.4} & 99.2 & \underline{99.8} & \textbf{93.7} & 97.5 & \textbf{94.9} & \underline{94.8} & 77.0 & \textbf{85.6} & \textbf{93.3} \\
                              & PA & \textbf{100.0} & \textbf{99.8} & \underline{72.2} & \textbf{100.0} & \underline{98.0} & \underline{99.3} & \underline{85.2} & \underline{98.2} & \underline{86.0} & \textbf{99.6} & \textbf{98.2} & \textbf{85.8} & 53.1 & \underline{62.6} & \textbf{88.4} \\
                              & FGA & \textbf{100.0} & \textbf{100.0} & \textbf{90.0} & \textbf{96.8} & \textbf{100.0} & 80.5 & 92.0 & \underline{96.2} & 88.9 & \textbf{90.2} & \textbf{87.9} & \underline{92.4} & \underline{81.8} & \textbf{83.0} & \underline{91.4} \\
                              & FTA & \textbf{100.0} & \textbf{86.2} & \textbf{90.0} & \textbf{94.6} & \underline{93.8} & \textbf{81.6} & \textbf{87.4} & \textbf{88.0} & \underline{74.0} & \textbf{76.4} & \textbf{80.1} & \underline{73.5} & \textbf{57.0} & \textbf{59.9} & \textbf{81.6} \\

\bottomrule

\end{NiceTabular}%
}
\end{table*}%

\subsection{RQ1: Effectiveness}
\noindent \textbf{Setup:} In the first research question (RQ), we evaluate the accuracy of \nm by comparing it with state-of-the-art log parsers, as accuracy is the most critical factor for log parsers.
For comprehensiveness, we compare \nm with both unsupervised parsers, which do not require human-labelled log templates, and label-required semantic-based log parsers, which rely on labelled log templates to support training, fine-tuning, or in-context learning.
Table.~\ref{tab:RQ1} presents the overall results on four metrics for \nm compared to baseline methods.
The best results for each metric on each dataset are highlighted in \textbf{bold}, while the second-best results are \underline{underlined}. 
If a specific parser could not complete the parsing process within a reasonable timeframe (\eg 12 hours), as per previous work~\cite{zhu2019tools,khan2022guidelines,jiang2023llmparser,jiang2023large}, we denote the score as “-”.

\noindent \textbf{Results:} 
Overall, we observe that \nm achieves the best average grouping accuracy (GA), parsing accuracy (PA), and F1 score of template accuracy (FTA), as well as the second-best F1 score of grouping accuracy (FGA).
Additionally, \nm demonstrates high accuracy across all 14 datasets, showcasing its robustness in parsing log data from diverse systems.
For example, \nm achieves the highest FTA on 11 out of 14 datasets and the second-highest FTA on the remaining 3 datasets.

Among unsupervised log parsers, it is evident that \nm significantly outperforms the other four baselines across all four evaluation metrics.
Among the syntax-based parsers (\ie AEL, Drain, and Brain), Brain achieves the highest scores in template-level metrics, with an average FGA of 71.8\% and an average FTA of 35.4\%.
However, LUNAR exhibits a substantially higher average FGA of 91.4\% and FTA of 81.6\%, surpassing Brain by 19.6\% and 46.2\%, respectively.
This is primarily due to the limitations of syntax-based parsers, which rely solely on manually crafted rules and thus struggle with complex and diverse log data. 
Furthermore, leveraging the extensive pre-trained knowledge of LLMs, LILAC w/o ICL demonstrates superior performance than syntax-based methods in PA, FGA, and FTA. 
However, our label-free LLM-based parser, \nm, significantly outperforms LILAC w/o ICL, with an  average improvement of 15.2\% in FGA and 26.9\% in FTA, respectively.
These substantial improvements underscore the effectiveness of \nm in harnessing the zero-shot capabilities of LLMs without labelled examples.

Compared with supervised baseline parsers, \nm demonstrate comparable performance to existing state-of-the-art parser LILAC. 
Specifically, compared to LILAC, \nm has achieved superior performance in average GA (93.3\%), PA (88.4\%), and FTA (81.6\%), while showing a slightly lower average FGA score (91.4\%). LILAC retrieves similar logs and labelled templates to provide demonstrations during parsing, which are often inaccessible in real-world systems. However, \nm achieves a comparable performance to the state-of-the-art while does not need labelled templates, which is more robust and generalizable in practice.   
Additionally, compared to UniParser and LogPPT, \nm has achieved substantial improvements with an average improvements of 30.2\% and 28.7\% in all four metrics. This result again highlights the advantages of \nm, which is not also effective, but also requires no labelled data for training. 
The comparable performance of \nm indicates its potential to be applied in real-world production systems, which can address the label dependency problems of semantic-based log parsers as mentioned in Section~\ref{sec:limitation}.

\subsection{RQ2: Impact of different settings}

\subsubsection{Designs}

In this research question, we first assess the individual contributions of the designs in the two components of \nm: the hierarchical log sharder and the two-stage LCU selector.
To accomplish this, we have implemented several variants of \nm by either removing or replacing the designs in these components.
Specifically, for the hierarchical log sharder, we created two variants by removing the respective clustering stages.
Additionally, for the two-stage LCU selector, we replaced the hybrid LCU ranking algorithm with various alternatives: random selection, simple selection based on minimum or maximum similarity, and selection of consecutive log messages for querying the LLMs.
To mitigate the impact of randomness, we also repeated experiment five times and calculated the mean scores as final results.

The average metric scores across all datasets are presented in Table \ref{tab: RQ2-ablation}.
When the log length-based clustering is removed from the log sharder, the end-to-end accuracy of \nm drops from 93.4\% to 87.7\% and from 88.4\% to 82.3\%, respectively.
Similar decreases are observed for the variant without top-k tokens clustering.
These results indicate that both stages within the hierarchical log sharder significantly contribute to the overall performance of \nm.
On the other hand, for the four variants related to the two-stage LCU selector, all four metrics show decreases across the board.
For example, replacing the hybrid LCU ranking algorithm with random selection results in average GA and PA scores dropping by 7.4\% and 6.6\%, respectively.
Notably, selecting the LCU based on maximum similarities among log messages has the most detrimental impact on \nm's performance, resulting in a 10.2\% decrease in GA and an 8.7\% decrease in PA.
This is primarily because selecting only the most similar log messages as an LCU for prompting prevents the LLM from accurately identifying templates and parameters by contrasting the tokens of these log messages.
These results demonstrate that our proposed two-stage LCU selector is effective in selecting LCUs, thereby enabling LLMs to accurately parse log messages.

\begin{table}[]
\captionsetup{justification=centering}
\centering
\caption{Ablation study of components in \nm(\%)}
\vspace{-12pt}
\label{tab: RQ2-ablation}
 \resizebox{0.48\textwidth}{!}{%
\begin{NiceTabular}{lcccc}
\toprule
\textbf{Metrics} & \textbf{GA} & \textbf{PA} & \textbf{FGA} & \textbf{FTA} \\
\hline
\nm & 93.4 & 88.4 & 91.4 & 81.6 \\
\midrule
\rowcolor{grey}\multicolumn{5}{c}{\textbf{Variants w.r.t Log Sharder}} \\
\midrule
w/o log length & 87.7 ($\downarrow 5.7\%$) & 82.3 ($\downarrow 6.1\%$) & 88.6 ($\downarrow 2.8\%$)& 80.4 ($\downarrow 1.2\%$)\\
w/o top-k tokens &  91.6 ($\downarrow 1.8\%$) & 84.3 ($\downarrow 4.1\%$) & 90.8 ($\downarrow 0.6\%$) & 81.0 ($\downarrow 0.6\%$)\\
\midrule
\rowcolor{grey}\multicolumn{5}{c}{\textbf{Variants w.r.t LCU Selector}} \\
\midrule
w/ random selection & 86.0 ($\downarrow 7.4\%$) & 81.8 ($\downarrow 6.6\%$) & 88.7 ($\downarrow 2.7\%$)& 79.8 ($\downarrow 1.8\%$)\\
w/ minimum similarity & 88.8 ($\downarrow 4.6\%$) & 83.9 ($\downarrow 4.5\%$) & 89.5 ($\downarrow 1.9\%$)&  80.2 ($\downarrow 1.4\%$)\\
w/ maximum similarity & 83.2 ($\downarrow 10.2\%$) & 79.7 ($\downarrow 8.7\%$) & 87.2 ($\downarrow 4.2\%$)& 78.6 ($\downarrow 3.0\%$)\\
w/ consecutive selection& 83.5 ($\downarrow 9.9\%$) & 80.2 ($\downarrow 8.2\%$) & 88.5 ($\downarrow 2.9\%$)& 80.0 ($\downarrow 1.6\%$)\\
\bottomrule
\end{NiceTabular}
\vspace{-12pt}
}
\vspace{-12pt}
\end{table}

\subsubsection{Configurations}

\begin{figure}[t]
    \centering
    \includegraphics[width=\columnwidth]{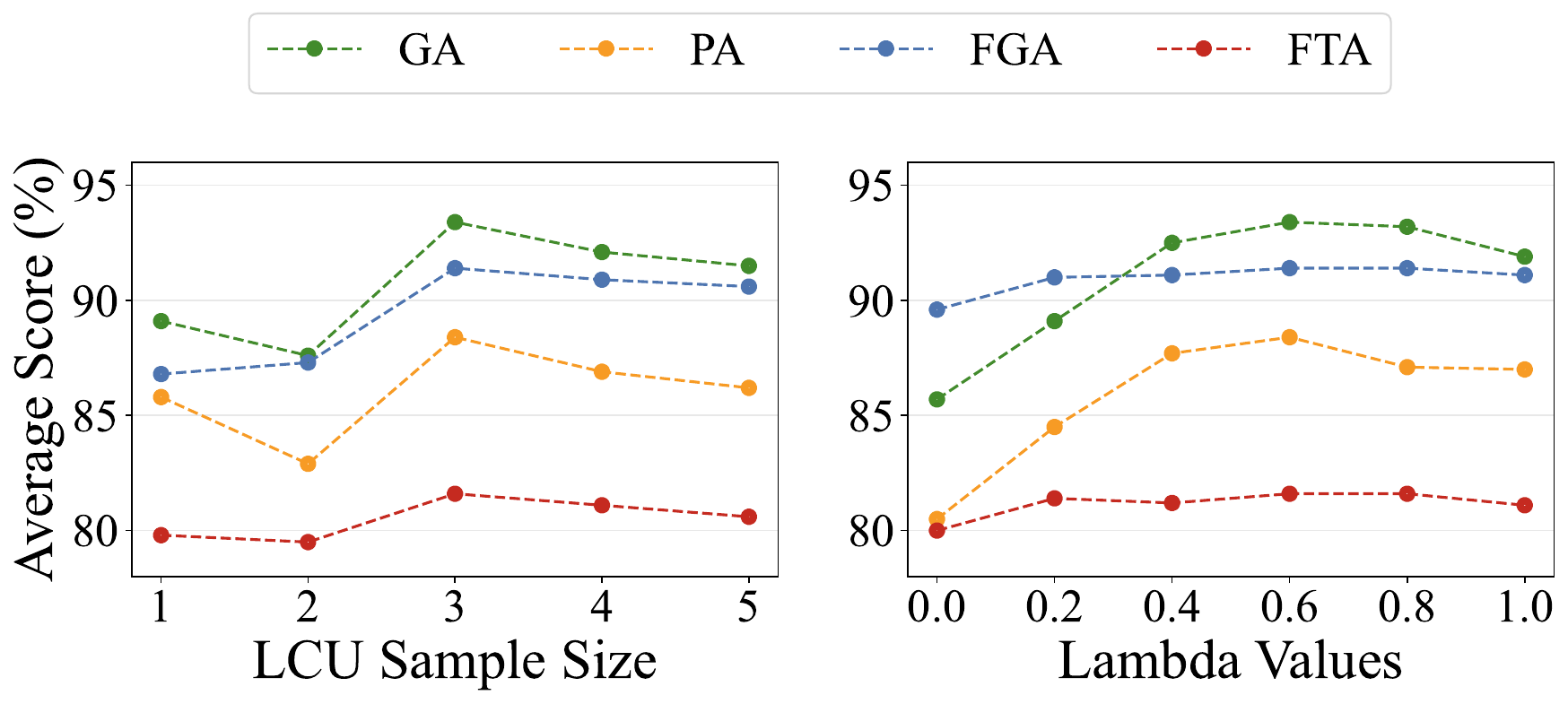}
    \caption{Sensitivity study of configurations in \nm.}
    \label{fig:RQ2_sensitivity}
\end{figure}

In addition to the two key designs of \nm mentioned above, we have identified two configurations that could significantly affect \nm's performance: the LCU sample size and the $\lambda$ values for LCU nomination.
These configurations directly determine the LCUs for each query provided to the LLMs, and thus, they may significantly impact the accuracy of the LLMs' parsed results.
In this part, we explore different settings for the LCU sample size and the $\lambda$ values to evaluate their impact on \nm's performance across all four evaluation metrics.
Specifically, we first chose the default $\lambda$ value (\ie 0.5) and varied the LCU sample sizes from 1 to 5. Additionally, we fixed the default LCU sample size (\ie 3) and varied the $\lambda$ value from 0.0 to 1.0.
The average metric scores under different settings across all datasets are presented in Figure~\ref{fig:RQ2_sensitivity}.

Based on the results shown in the left sub-figure of Figure~\ref{fig:RQ2_sensitivity}, we observe that the performance of \nm remains consistently high across different LCU sample sizes.
Notably, when the LCU sample size is set to 3, all four metrics achieve their highest scores.
Conversely, smaller sample sizes result in lower accuracy.
For instance, with a sample size of 1, the average GA and PA scores drop to approximately 89\% and 86\%, respectively, which is 4.5\% and 2.5\% lower than the scores achieved with a sample size of 3.
This suggests that the absence of contrastive log messages can impair the parsing ability of LLMs.
Interestingly, when the sample size is 2, the accuracy decreases further compared to a sample size of 1.
This occurs because with only 2 contrastive log messages, the LLM tends to overfit to the differences between them, neglecting potential variable tokens in the log messages.
On the other hand, increasing the LCU sample size to 4 or 5 results in a downward trend in accuracy.
For example, the FTA score decreases from 81.6\% to 80.6\%.
This indicates that larger LCU sample sizes may introduce additional noise and thus do not enhance the accuracy of \nm.
Therefore, we have set the default LCU sample size to 3 in our experiments.

The $\lambda$ values for LCU nomination take into account both variability and commonality scores when selecting LCUs for prompting.
A higher $\lambda$ value places more emphasis on variability, while a lower $\lambda$ value prioritizes commonality.
As shown in Figure \ref{fig:RQ2_sensitivity}, the optimal average accuracy is achieved when the $\lambda$ value is set to 0.6, indicating a balanced consideration of variability and commonality in log messages within the LCU.
When the $\lambda$ value is low, such as 0.0 or 0.2, all metrics are significantly lower compared to a value of 0.6, underscoring the importance of variability in log messages within LCUs.
For instance, at a $\lambda$ value of 0.0, the average GA and PA scores are both 8\% lower than at 0.6.
Conversely, performance also declines when the $\lambda$ value is too high.
For example, both GA and PA scores drop by 1.5\% when the $\lambda$ value increases from 0.6 to 1.0.
These results illustrate that both variability and commonality scores effectively filters log messages belonging to the same template within the LCUs, which enable LLMs to accurately parse the log template.

\subsection{RQ3: Efficiency}\label{sec:experiment-rq3}

\begin{figure}[t]
    \centering
    \includegraphics[width=\columnwidth]{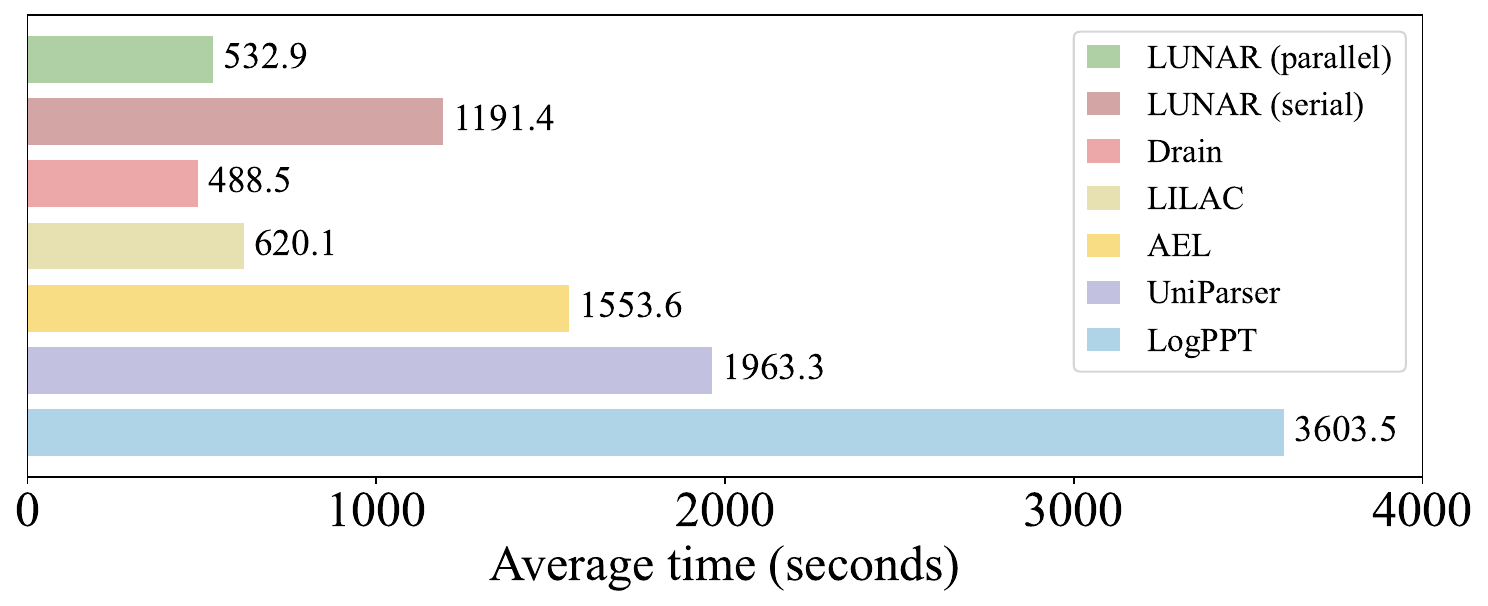}
    \caption{Efficiency of \nm and baselines (second).}
    \label{fig:RQ3_efficiency}
\end{figure}

\noindent \textbf{Setup:}
Efficiency is an essential factor for log parsers in real-world usage, given the substantial volume of logs produced~\cite{zhu2019tools,wang2022spine}.
In this RQ, we evaluate the efficiency of \nm and all other baseline parsers by apply them in the large-scale datasets within Logpub, which contains average 3.6 million log messages per dataset.
Specifically, we recorded the execution times for all log parsers in parsing all log datasets within Logpub, and then compute the average parsing time across all datasets.
Since \nm is designed to enable parallelism for processing log buckets, we compute the average parsing time for both serial mode and parallel mode of \nm.
The efficiency results are demonstrated in the Figure~\ref{fig:RQ3_efficiency}.

\noindent \textbf{Results:}
The results demonstrate that \nm achieves higher efficiency compared to the most efficient semantic-based log parser, LILAC, and comparable efficiency to the most efficient syntax-based log parser, Drain. Specifically, the average parsing time for 3.6 million log messages in serial mode using \nm is 1191.4 seconds.
In practice, we can leverage the parallel mode of \nm to achieve better efficiency, reducing the parsing time to just 532.9 seconds.
For the most efficient syntax-based parser, Drain, which only takes an average of 488.5 seconds to parse each dataset, \nm achieve comparable efficiency, only slower by 9\%.
In contrast, the most efficient semantic-based parser, LILAC, takes an average of 620.1 seconds,
which is 16.4\% slower than \nm.
Notably, even with GPU acceleration, other semantic-based parsers such as UniParser and LogPPT require significantly more time to parse large-scale log data. \nm outperforms them by 3.68 times and 6.75 times in efficiency, respectively.
These results indicate that \nm is efficient in parsing large-scale log data, making it suitable for application in real-world production systems.

\section{THREATS TO VALIDITY}

We have identified the following major threats to validity:

\noindent
\textbf{Data leakage}
Given that large language models (LLMs) are trained on extensive datasets, one potential risk is data leakage.
Specifically, the LLM used in \nm might have been trained on open-source log datasets, which could result in memorizing ground-truth templates rather than performing genuine inference.
However, experimental results indicate that the performance of merely using LLMs (LILAC w/o ICL) is significantly lower compared to \nm, suggesting a low likelihood of direct memorization.
Additionally, \nm utilizes the same LLM, \ie \textit{gpt-turbo-3.5-0613} model for experiments, aligning with previous work, LILAC.

\noindent
\textbf{Randomness}
Randomness can influence the performance of \nm and other baseline methods.
To mitigate this issue, we minimized the randomness of the LLM by setting the temperature to 0, ensuring consistent outputs for the same input text.
Additionally, we conducted each experiment five times for every experimental setting and used the average of these results as the final outcome.

\noindent
\textbf{Implementation and settings}
To mitigate the bias of implementation and settings, in our evaluation, we compared our \nm with state-of-the-art approaches within the same evaluation framework.
We adopted the implementations from their replication packages and benchmarks, using the parameters and settings (\eg number of log templates and similarity threshold) optimized by previous work~\cite{zhu2019tools,jiang2023large}.
Moreover, the results of the baseline approaches are consistent with the best results in recent benchmarks.

\section{Related Works}

Log parsing is a critical preliminary step for various log analysis tasks~\cite{khan2022guidelines}, including anomaly detection and root cause analysis.
Therefore, numerous efforts have been made to achieve accurate and efficient log parsing~\cite{nagappan2010abstracting,vaarandi2015logcluster,dai2020logram,shima2016LenMa,tang2011logsig,makanju2009clustering,messaoudi2018search,mizutani2013SHISO,du2016spell,yu2023brain}.
These log parsers can be categorized into two types: unsupervised syntax-based and supervised semantic log parsers.
Unsupervised syntax-based log parsers leverage predefined rules or heuristics to extract the constant parts of log messages as log templates.
For instance, SLCT~\cite{vaarandi2003data} was the first approach to use token frequencies to determine log templates and parameters.
LogMine~\cite{hamooni2016logmine} employs a bottom-up clustering algorithm to segment log messages based on customized similarity measures, and extract log templates for each cluster.
Furthermore, Drain~\cite{he2017drain} utilizes a fixed-depth prefix tree structure to effectively extract commonly occurring templates based on specific heuristics (\ie prefix tokens and log length).
These syntax-based log parsers do not rely on manually labelled examples.
However, their parsing accuracy can significantly decline when log data do not conform to predefined rules.

On the other hand, semantic-based log parsers utilize neural networks~\cite{li2023did,liu2022uniparser} or language models~\cite{le2023log} to identify log templates and parameters by understanding the semantics of log messages.
They require human-labelled log templates to train or tune the models by learning the semantics and patterns in the labelled log data.
UniParser~\cite{liu2022uniparser} is one of the pioneering work in parsing logs with a focus on their semantic meaning.
It integrates a BiLSTM-based semantics miner with a joint parser to identify log templates.
Additionally, LogPPT~\cite{le2023log} proposed identifying log templates and parameters using prompt-based few-shot learning, based on the RoBERTa model.
Recently, with the rise of large language models (LLMs), a series of LLM-based log parsers have been developed to achieve more effective log parsing.
These LLM-based log parsers leverage fine-tuning~\cite{liu2023logprompt} or in-context learning~\cite{xu2023prompting,jiang2023llmparser} to specialize LLMs for log parsing tasks, thereby achieving remarkable performance.
However, these semantic-based log parsers heavily rely on labelled data, which limits their ability to generalize to different types of log data or evolving log data~\cite{jiang2023large}.
In contrast, our proposed unsupervised LLM-based log parser, \nm, does not require labelled examples, allowing it to generalize to diverse and evolving log data.

\section{Conclusion}
In this work, we propose an LLM-based unsupervised log parser named \nm, which leverages log contrastive units (LCUs) to facilitate effective comparisons by the LLM.
To efficiently identify effective LCUs from large-scale log data, \nm employs a hierarchical sharder to divide logs into buckets, thereby reducing sampling overhead and enabling parallel computation.
Additionally, \nm incorporates a hybrid LCU selector that jointly measures the commonality and variability of LCUs, which is crucial for prompting LLMs.
Furthermore, \nm introduces an improved prompt format to guide LLMs in a zero-shot setting.
Experimental results on 14 large-scale public datasets demonstrate that \nm achieves high parsing accuracy, significantly outperforming other unsupervised parsers and being comparable to state-of-the-art parsers that require labelled data.
In terms of efficiency, \nm is also comparable to the fastest baseline parser, highlighting its potential for application in real-world systems.

\balance
\bibliographystyle{ACM-Reference-Format}
\bibliography{reference}

\end{document}